\documentclass[letterpaper, 12pt]{spieman}  
\usepackage{amsmath,amsfonts,amssymb}
\usepackage{graphicx}
\usepackage{setspace}
\usepackage{tocloft}
\usepackage{lineno}

\usepackage[top=1in, left=0.95in, right=0.95in, bottom=1.2in]{geometry}

\title{MNTES: Modeling Nonlinearity of TES detectors for Enhanced Cosmic Microwave Background measurements with \emph{LiteBIRD}}

\author[a,b]{Tijmen de Haan}
\author[ ]{for the LiteBIRD Collaboration}
\affil[a]{Institute of Particle and Nuclear Studies (IPNS), High Energy Accelerator Research Organization (KEK), Tsukuba, Ibaraki 305-0801, Japan}
\affil[b]{International Center for Quantum-field Measurement Systems for Studies of the Universe and Particles (QUP-WPI), High Energy Accelerator Research Organization (KEK), Tsukuba, Ibaraki 305-0801, Japan}

\cftpagenumbersoff{figure}
\cftpagenumbersoff{table} 
\begin{document} 
\maketitle

\begin{abstract}
Traditional methods of converting electronic readout counts to optical power incident on Transition Edge Sensors (TES) for Cosmic Microwave Background (CMB) observations involve a linear approximation. For the upcoming \emph{LiteBIRD} CMB satellite, strict nonlinearity requirements must be met to prevent contamination of the science band at $4f_\mathrm{HWP}$ by the $2f_\mathrm{HWP}$ signal, which arises from differential transmission or emissivity related to the half-wave plate's rotation rate $f_\mathrm{HWP}$. These constraints cannot be met using hardware solutions alone and therefore require a form of nonlinearity correction. We present MNTES, a novel physics-based, nonlinear calibration technique. This method leverages our physical understanding of the TES power balance equation, accounts for imperfect voltage bias by casting the bias network as its Thévenin equivalent, and can incorporate external information such as time-varying magnetic fields and focal plane temperature variations. The detector-specific parameters of MNTES will be measured during the ground calibration campaign prior to the \emph{LiteBIRD} launch, yielding conversion functions that can take raw time-ordered data and output the reconstructed incident optical power. MNTES will allow us to achieve \emph{LiteBIRD}’s goal of measuring the primordial tensor fluctuation spectrum to $\delta r < 0.001$.
\end{abstract}


{\noindent \footnotesize\textbf{*}Tijmen de Haan,  \linkable{tijmen.dehaan@gmail.com} }



\section{Introduction}
\label{sect:intro}

The Cosmic Microwave Background (CMB) has proven to be the most successful cosmological probe for understanding the origins and evolution of the universe. While low-resolution mapping of CMB temperature anisotropies has largely been completed by the \emph{WMAP} \cite{bennett_nine-year_2013} and \emph{Planck} \cite{aghanim_planck_2020} satellites, frontier efforts now focus on high-resolution CMB temperature mapping \cite{sobrin_design_2022, abazajian_snowmass_2022} and detailed characterization of CMB polarization. Precise measurements of CMB polarization are expected to yield new insights into cosmic inflation, the composition of the universe, and the formation of large-scale structure. Of particular interest is the potential to detect the faint imprint of primordial gravitational waves, quantified by the tensor-to-scalar ratio $r$ \cite{kamionkowski_quest_2016}.

The \emph{Lite} satellite for the studies of \emph{B}-mode polarization and \emph{I}nflation from cosmic background \emph{R}adiation \emph{D}etection (\emph{LiteBIRD}) is a JAXA-led mission scheduled for launch in 2032 \cite{litebird_collaboration_probing_2023}. \emph{LiteBIRD} aims to map the polarization of the CMB over the full sky with unprecedented sensitivity, targeting a measurement of the tensor-to-scalar ratio with precision $\delta r < 0.001$. Achieving this ambitious goal requires exquisite control of systematic effects and sources of spurious polarization.
\emph{LiteBIRD} will employ arrays of transition-edge sensor (TES) bolometers \cite{westbrook_detector_2021} read out using digital frequency multiplexing (DfMux) \cite{dobbs_frequency_2012} to detect the CMB at millimeter wavelengths. While TES bolometers provide state-of-the-art sensitivity, they exhibit intrinsic nonlinearity in their response due to the nature of the superconducting transition \cite{ghigna_modelling_2023}. Non-ideal voltage biasing introduces additional nonlinearity.

TES nonlinearity poses several challenges for \emph{LiteBIRD}'s science goals. Most significantly, it can up-convert signals from the continuously rotating half-wave plate at $2f_\mathrm{HWP}$ to the science band at $4f_\mathrm{HWP}$, where $f_\mathrm{HWP}$ is the rotation frequency. Nonlinearity can also distort the shapes of bright astrophysical sources, bias the calibration derived from the CMB temperature dipole, and cause gain variations from drifts in optical loading, focal plane temperature, or magnetic field environment. Tight control of these systematic effects is essential for achieving \emph{LiteBIRD}'s target sensitivity.

Previous attempts to correct for nonlinearity have typically relied on empirical fitting functions\cite{den_herder_performance_2016, villa_planck_2010}. In this paper, we introduce \textbf{M}odeling \textbf{N}onlinearity of \textbf{TES} detectors (MNTES), a novel method to characterize and correct for nonlinearity in \emph{LiteBIRD}. MNTES employs a physics-based model of the TES power balance and the effective readout circuit to enable a more accurate reconstruction of the incident optical power compared to standard linear methods. 

The rest of this paper is organized as follows. Section~\ref{sect:need_for_mntes} details the potential impact of nonlinearity on \emph{LiteBIRD}'s science goals. Section~\ref{sect:placeholder_model} describes our model for a TES bolometer under DfMux readout, including the effects of non-ideal voltage bias and temperature and current-dependent resistance. Section~\ref{sect:mntes_reconstruction_method} presents the MNTES nonlinear reconstruction method, and we summarize in Section~\ref{sect:conclusion}.

\section{Effect of Nonlinearity in \emph{LiteBIRD}}
\label{sect:need_for_mntes}

In this section, we describe several ways in which nonlinearity in the TES bolometers can degrade \emph{LiteBIRD}'s science return if left uncorrected. These effects include the up-conversion of half-wave plate synchronous signals, saturation on bright astrophysical sources, bias in the CMB dipole calibration, and gain variations from drifts in the detector operating point. Understanding and mitigating these nonlinearity effects is crucial for achieving \emph{LiteBIRD}'s target sensitivity.

\subsection{Half-Wave Plate Systematics}
\label{subsect:hwp_systematics}

\emph{LiteBIRD} will employ a continuously rotating half-wave plate (HWP) \cite{klein_cryogenic_2011} as the first optical element in order to modulate the incident polarization. The HWP has a non-negligible differential transmission/emissivity of $\sim 1\%$ arising from differences in the reflection and absorption coefficients along the ordinary and extraordinary axes of the birefringent material. These effects generate a signal at the second harmonic of the HWP rotation frequency, $2f_\mathrm{HWP}$. 

Non-zero absorption in the HWP, vibrational coupling, and magnetic coupling provide additional contributions to the $2f_\mathrm{HWP}$ signal. If the detector response were perfectly linear, this $2f_\mathrm{HWP}$ signal would lie outside \emph{LiteBIRD}'s science band. However, nonlinearity can up-convert the $2f_\mathrm{HWP}$ signal to $4f_\mathrm{HWP}$, where it contaminates the sky polarization signal. Furthermore, any $1/f$ noise on the HWP synchronous signal, for example from temperature fluctuations of the HWP, will also be up-converted to the science band.

We present a detailed analysis of this $2f_\mathrm{HWP} \rightarrow 4f_\mathrm{HWP}$ conversion in the companion paper by Micheli et al. (2024, in preparation) \cite{micheli_systematic_2024}, where we show that the nonlinearity must be controlled to $<0.014~\mathrm{pW}^{-1}$ to meet \emph{LiteBIRD}'s systematic error budget.

\subsection{Saturation on Bright Sources}
\label{subsect:saturation_jupiter}

Although \emph{LiteBIRD}'s detectors are optimized for measuring the faint CMB polarization signals, they will also encounter much brighter sources during the full-sky survey. The planet Jupiter is expected to be the brightest compact source, contributing up to $\sim 0.8~\mathrm{pW}$ of optical power in \emph{LiteBIRD}'s highest frequency band at 448 GHz \cite{litebird_collaboration_probing_2023}. The Galactic plane will also appear as a bright extended source.

Nonlinearity will distort the shapes of these bright sources in the maps, reducing our ability to use them for calibrating the temperature response and beam properties. Modeling and correcting for nonlinearity is necessary to extend the usable dynamic range of the detectors for these bright sources.

\subsection{Effects of Nonlinearity on the CMB Dipole}  
\label{subsect:non_linearity_dipole}

The CMB dipole, arising from the motion of the solar system with respect to the CMB rest frame, is a key calibrator for space-based CMB missions. The CMB dipole has an amplitude of $3.36~\mathrm{mK}_\mathrm{CMB}$ \cite{aghanim_planck_2020}. While this is a relatively small signal, nonlinearity can still cause a detectable shift in the overall gain calibration derived from the dipole. Therefore, nonlinearity mitigation with MNTES is desirable.

\subsection{Gain Drifts}
\label{subsect:gain_drifts}

TES bolometers are sensitive to the operating point set by parameters such as the optical loading, bath temperature, and magnetic field. Slow drifts in these parameters can shift the bias point, changing the local detector responsivity. Although thermal mass and magnetic shielding can suppress low-frequency fluctuations to some extent, nonlinearity corrections are needed to remove residual gain drifts and stabilize the calibration over long time scales.

\subsection{ADC Nonlinearity}
\label{subsect:planck_adc_nonlinearity}

We note that nonlinearity in the analog-to-digital converters (ADCs) was a significant systematic effect for the \emph{Planck} mission \cite{aghanim_planck_2014, adam_planck_2016}. However, this is much less of a concern for \emph{LiteBIRD} due to the different readout architecture. In contrast to \emph{Planck}'s individual ADCs for each detector, the \emph{LiteBIRD} DfMux system combines $\sim 60$ detector signals into a single ADC \cite{montgomery_performance_2022}. The ADC nonlinearity is effectively randomized across many detectors and appears as an additional white noise term that is negligible compared to other noise sources.

\section{TES Detector and DfMux Readout Model}
\label{sect:placeholder_model}

A transition-edge sensor (TES) consists of a superconducting thin film voltage-biased in its resistive transition. When additional optical power is absorbed by the TES island, it heats up, changing its resistance. This change is largely counteracted by a corresponding decrease in electrical power dissipation, which depends on resistance. The strength of this \textit{electrothermal feedback} \cite{irwin_application_1995} loop is characterized by a loop gain, in analogy to the open-loop gain of an amplifier under feedback:
\begin{equation*}
    \mathcal{L}\left( \delta P_\mathrm{opt} + \delta P_\mathrm{el} \right) = - \delta P_\mathrm{el}
\end{equation*}
where $\mathcal{L}$ is the loop gain, $\delta P_\mathrm{opt}$ is the perturbation in the absorbed optical power, and $\delta P_\mathrm{el}$ is the perturbation in the electrical power dissipated in the TES. 

\emph{LiteBIRD} will read out its array of $\sim5000$ TES bolometers using digital frequency multiplexing (DfMux), a system similar to those used for Simons Array (SA) \cite{barron_integrated_2021} and SPT-3G \cite{montgomery_performance_2022}. The readout circuit contains non-idealities in the form of static parasitic impedances, which are relatively well-understood \cite{montgomery_performance_2022, zhou_method_2024} thanks to the long heritage of this technology. However, we emphasize that a detailed understanding of these non-idealities is not necessary for our nonlinearity modeling approach, MNTES. Instead, we rely on the AC version of Thévenin's theorem, which states that any linear circuit with multiple impedances, voltage sources, and current sources can be represented by a Thévenin-equivalent circuit consisting of an ideal voltage source with effective voltage $V_\mathrm{Th\acute{e}v}$ in series with an effective impedance $z_\mathrm{Th\acute{e}v}$. We adopt this Thévenin-equivalent circuit parameterization, with $V_\mathrm{Th\acute{e}v}$ and $z_\mathrm{Th\acute{e}v}$ to be determined during the on-ground calibration campaign.

In equilibrium, the TES is governed by the power balance equation \cite{mather_bolometer_1982, irwin_transition-edge_2005, lee_voltage-biased_1998}:
\begin{equation}
P_\mathrm{opt} + \frac{V_\mathrm{Th\acute{e}v}^2 R(T, I)}{\left| R(T,I) + z_\mathrm{Th\acute{e}v} \right|^2}  = K \left( T^n - T_\mathrm{bath}^n \right) \ ,
\label{eq:power_balance}
\end{equation}
where $T$ is the TES temperature, $P_\mathrm{opt}$ is the absorbed optical power, $K$ is a thermal conductance parameter, $T_\mathrm{bath}$ is the bath temperature ($\sim 100~\mathrm{mK}$ for \emph{LiteBIRD}), and $n$ is an exponent that depends on the thermal transport mechanism. The function $R(T,I)$ describes the TES resistance as a function of both temperature and current, often modeled using a nonlinear transition model such as a resistively-shunted junction, two-fluid, or vortex street model \cite{gottardi_review_2021}. In this work, we treat $R(T, I)$ as an unknown differentiable function. 

To calculate the loop gain of electrothermal feedback $\mathcal{L}$ and the small-signal responsivity $S = dI/dP$, we start from the power balance equation (Eq. \ref{eq:power_balance}) and perform a first-order Taylor expansion around an operating point $(T, I)$, assuming the small-signal fluctuations are sufficiently slow to maintain (quasi-)equilibrium:

\begin{equation*}
\delta P_\mathrm{opt} + \frac{V_\mathrm{Th\acute{e}v}^2}{\left| R + z_\mathrm{Th\acute{e}v} \right|^2} \delta R - \frac{2 V_\mathrm{Th\acute{e}v}^2 R (R + \Re{(z_\mathrm{th\acute{e}v})})}{\left| R + z_\mathrm{Th\acute{e}v} \right|^4} \delta R = K n T^{n-1} \delta T \ .
\end{equation*}
assuming that $R>\Re{(z_\mathrm{Th\acute{e}v})}$ with $\Re$ denoting the operation of taking the real part of a complex quantity. Using the power law expansion of the resistance $\delta R / R = \alpha \delta T / T + \beta \Re (\delta I / I)$ and Ohm's law $\delta V_\mathrm{Th\acute{e}v} = \left( R + z_\mathrm{Th\acute{e}v} \right) \delta I + I \delta R$, we obtain
\begin{equation}
  \boxed{\mathcal{L} = \frac{\alpha V_\mathrm{Th\acute{e}v}^2}{K n T^n R} \frac{R^2 \left( R^2 - \left|  z_\mathrm{Th\acute{e}v} \right |^2 \right)}{\left| R + z_\mathrm{Th\acute{e}v} \right|^2 \left( \left| R + z_\mathrm{Th\acute{e}v} \right|^2 + \beta R (R +  \Re{(z_\mathrm{Th\acute{e}v})}) \right) }}    
\end{equation}
for the loop gain of the electrothermal feedback. The responsivity is defined as the ratio of the measured current fluctuation to the incident optical power fluctuation
\begin{equation}
    \boxed{ S \equiv \frac{\delta I}{\delta P_\mathrm{opt}} = - \frac{\sqrt{2}}{V_\mathrm{Th\acute{e}v}} \frac{\mathcal{L}}{\mathcal{L} + 1} \left( 1 + 2 z_\mathrm{Th\acute{e}v}^\star \frac{R + \Re{(z_\mathrm{Th\acute{e}v})}}{R^2 - \left| z_\mathrm{Th\acute{e}v} \right|^2 } \right) } \ .
    \label{eq:responsivity_full}
\end{equation} 
Note that the complex conjugate of $z_\mathrm{Th\acute{e}v}$ appears in the responsivity, meaning that when there is non-negligible effective series inductance, the TES responds at a phase different from that of the voltage bias\cite{farias_understanding_2024}.

\subsection{Fiducial Parameter Values}
\label{subsect:qualitative_parameter_dependence}

To illustrate the qualitative behavior of a representative \emph{LiteBIRD} TES bolometer, we consider a placeholder transition model $R(T,I)$ that follows an arctangent function of temperature that asymptotes to the normal resistance $R_\mathrm{normal}$ and has a transition width parameterized by the interquartile range $\mathrm{IQR}$, with no explicit current dependence:
\begin{equation*}
    R(T) = \left(\frac{1}{2} + \frac{1}{\pi} \arctan\left(\frac{T - T_c}{{\mathrm{IQR}}/{2}}\right)\right) R_{\text{normal}}
\end{equation*}
In addition, we choose fiducial parameters, summarized in Table \ref{tab:fiducial_parameters}. Together these form a of a fiducial LiteBIRD TES and its readout that is broadly consistent with the expected properties of the \emph{LiteBIRD} detectors. However, we emphasize that this is a placeholder model used for illustration purposes only, and will be replaced with a more accurate TES model to be chosen after fabrication and characterization of the \emph{LiteBIRD} TES bolometers.


\begin{table}[ht]
\centering
\caption{Fiducial parameters for a representative \emph{LiteBIRD} TES bolometer.} 
\label{tab:fiducial_parameters}
\begin{tabular}{c c}
\hline
\hline
\textbf{Assumed Parameter} & \textbf{Value}\\
\hline
Loop gain & $\mathcal{L} = 10$ \\
Thévenin-equivalent impedance & $z_\mathrm{Th\acute{e}v} = 0.05+0.05j~\Omega$ \\
Normal resistance & $R_\mathrm{normal} = 1~\Omega$ \\ 
Fractional operating resistance & $R_\mathrm{frac} = 0.7$ \\
Transition temperature & $T_c = 180~\mathrm{mK}$ \\
Thermal bath temperature & $T_\mathrm{bath} = 100~\mathrm{mK}$ \\  
Thermal transport exponent & $n = 3.6$ \\
Optical loading (at TES) & $P_\mathrm{opt} = 0.5~\mathrm{pW}$ \\
Saturation power & $P_\mathrm{sat} = 2.5 P_\mathrm{opt}$ \\
Effective thermal time constant & $\tau = 3~\mathrm{ms}$ \\
\hline  
\hline
\textbf{Derived Parameter} & \textbf{Value}\\
\hline
Responsivity & $S=1.8 \times 10^6~\mathrm{A/W}$ \\
Transition width & $\mathrm{IQR} = 1.1~\mathrm{mK}$ \\
Logarithmic temperature dependence & $\alpha=77$ \\
Logarithmic current dependence & $\beta=0$ \\
Intrinsic thermal time constant & $\tau_0 = 33~\mathrm{ms}$ \\  
Thermal conductance & $\overline{G} = 15~\mathrm{pW/K}$ \\
Thévenin-equivalent voltage & $V_\mathrm{Th\acute{e}v} = 0.8~\mu\mathrm{V}$ \\
\hline
\end{tabular}
\end{table}

\subsection{Leading Order Nonlinearity}

Although MNTES performs a full nonlinear reconstruction of incident optical power, exploring the leading-order nonlinearity $\partial^2 I / \partial P_\mathrm{opt}^2$ can nevertheless provide valuable insights. For the fiducial parameters, this quantity has a value of $1.9 \times 10^{18}~\mathrm{A/W^2}$. Divided by the responsivity, we get $0.8~\mathrm{pW}^{-1}$. This can be compared directly to the requirement of $0.014~\mathrm{pW}^{-1}$ derived in the companion paper, highlighting the need for significant nonlinearity correction.

\begin{figure}
\begin{center}
\begin{tabular}{c}
\includegraphics[width=0.97\linewidth]{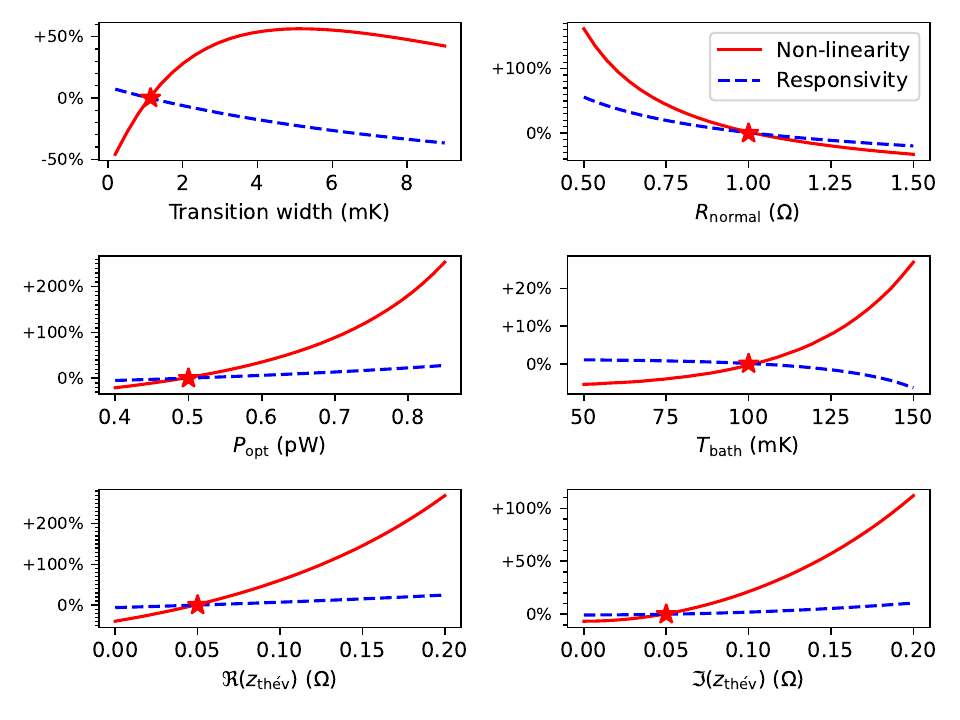}
\end{tabular}
\end{center}
\caption 
{ \label{fig:parameterdependence}
Dependence of responsivity and leading-order nonlinearity on assumed parameters. The stars show the fiducial parameter values from Table~\ref{tab:fiducial_parameters} and the curves show the effects of parameter variations on the responsivity and leading-order nonlinearity. The voltage bias is allowed to float in order to hold the fractional resistance fixed at the fiducial value of $R_\mathrm{frac}=0.7$.
}
\end{figure}

Figure~\ref{fig:parameterdependence} shows how both the responsivity itself and the leading-order nonlinearity depend on several key parameters. We adjust the voltage bias in order to keep the tuning point fixed to $R_\mathrm{frac}=0.7$, mimicking a tuned bolometer. We find some noteworthy behaviors:
\begin{itemize}
    \item If the actual optical loading is much higher than the design value, the TES must be biased at a lower voltage. While this raises the responsivity, it also leads to a steep rise in nonlinearity.
    \item The width of the superconducting transition has a relatively minor  effect on nonlinearity. This is because a very broad transition provides the TES with a large intrinsic dynamic range. On the other hand, a very narrow transition region results in strong electrothermal feedback, linearizing the TES. For the fiducial TES considered here, the nonlinearity is greatest if the transition width is $\sim5~\mathrm{mK}$.
    \item Introducing a series impedance spoils the voltage bias, boosting the responsivity of the TES. This can be desirable as it increases the signal size, decreasing the effective readout noise. However, achieving a 10\% boost in responsivity through this approach nearly doubles the nonlinearity, regardless of whether the series impedance is resistive or inductive.
\end{itemize} 

\section{MNTES Reconstruction Method}
\label{sect:mntes_reconstruction_method}

The goal of the MNTES procedure is to, at each time step, recover the input optical power incident on each detector from the value of the time-ordered data in digital readout counts.

During the on-ground calibration of \emph{LiteBIRD}, we will measure the physical parameters of the transition-edge sensors (TES) and readout system as part of the AIVC (Assembly, Integration, Verification, and Calibration) campaign. Specifically, for each bolometer, we will map out the measured complex current returned from Digital Active Nulling \cite{de_haan_improved_2012} $I_\mathrm{DAN}$ as a function of input optical power, local magnetic field, bias voltage, thermal bath temperature, and bias frequency. This calibration process can be described as measuring the function
\begin{equation}
    I_\mathrm{DAN}(P_\mathrm{opt}, B, V, T_\mathrm{bath}, \omega).
\end{equation}
This nonlinear mapping is crucial as it allows for the calibration of the TES bolometer response under varying conditions, essential for accurate power reconstruction in-flight.

The calibration involves systematically varying one parameter at a time to isolate different aspects of the bolometer response, allowing precise measurement of:
\begin{itemize}
    \item \textbf{Optical Response}, $I(P) = I_{\mathrm{DAN}}(P, B_0, V_0, T_0, \omega_0)$,
    \item \textbf{Magnetic Field Sensitivity}, $I(B) = I_{\mathrm{DAN}}(P_0, B, V_0, T_0, \omega_0)$,
    \item \textbf{Complex I-V Curve}, $I(V) = I_{\mathrm{DAN}}(P_0, B_0, V, T_0, \omega_0)$,
    \item \textbf{Impedance-Temperature Curve}, $R(T) + z_\mathrm{th\acute{e}v} = {V_0}/{I_{\mathrm{DAN}}(P_0, B_0, V_{\text{probe}}, T, \omega_0)}$,
    \item \textbf{Normal Network Analysis}, $Y(\omega) = {I_{\mathrm{DAN}}(P_0, B_0, V_{\text{probe}}, T>T_c, \omega)}/{V_0}$,
    \item \textbf{Superconducting Network Analysis}, $Y(\omega) = {I_{\mathrm{DAN}}(P_0, B_0, V_{\text{probe}}, T_0, \omega)}/{V_0}$.
\end{itemize}
The subscript $0$ denotes the fiducial value of each parameter used during normal detector operation. $V_\mathrm{probe}$ is a small probe voltage used to measure the impedance without significantly heating the TES island. The complex admittance $Y(\omega)$ measured both above and below the superconducting transition maps out the frequency-dependent transfer function due to the TES resistance and the readout circuit.

\begin{figure}
\begin{center}
\begin{tabular}{c}
\includegraphics[width=0.97\linewidth]{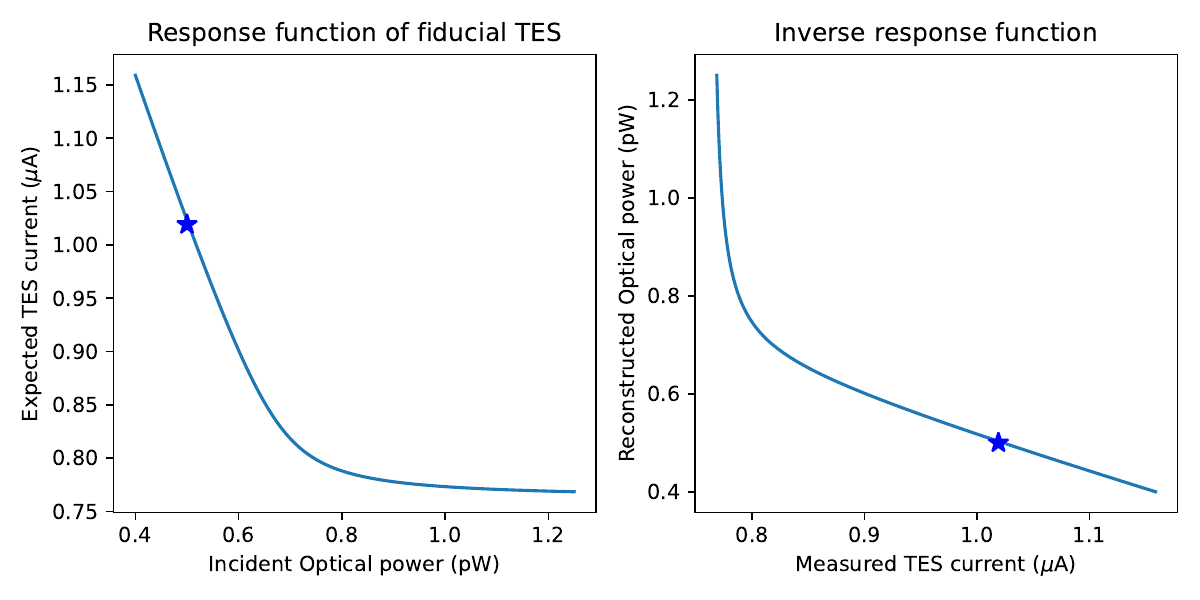}
\end{tabular}
\end{center}
\caption 
{ \label{fig:powerreconstruction}
\textit{Left:} MNTES response function for the fiducial TES specified in Table~\ref{tab:fiducial_parameters}. \textit{Right:} Inverse of the response function, which converts data collected from a TES with the fiducial parameters. This conversion process simultaneously calibrates and linearizes in-flight TES data, resulting in accurately reconstructed incident optical power.
}
\end{figure}

Once this characterization data has been obtained, the MNTES model can then be obtained for each TES separately. The overall normalization is obtained from the optical response measurement, while the information on readout parasitics and normal resistances is obtained from the network analyses. The transition is characterized by both the total impedance-temperature and I-V curves. This calibration dataset is then used to calculate the best-fit MNTES parameters from  Section~\ref{sect:placeholder_model}. This yields a per-detector physical model that ties the incident optical power to the measured readout counts with a relation such as the one shown in Figure~\ref{fig:powerreconstruction}. The model can be used to reconstruct incident optical power from in-flight data, even if the optical loading, bath temperature or external magnetic field are different from the on-ground calibration setup.

The end result is one-to-one conversion function as shown in Figure~\ref{fig:powerreconstruction} which takes the in-flight TES data samples and converts them individually to reconstructed incident optical power samples, effectively undoing the known sources of nonlinearity. This optical power data can then be fed into the mapmaking pipeline. 

\subsection{Numerical Implementation}

Evaluation of the MNTES model requires the use of a numerical solver for finding the point of power balance (Equation~\ref{eq:power_balance}). We use Brent's algorithm \cite{brent_algorithms_1973} but note that solving the power balance equation numerically for each sample in the entire LiteBIRD dataset is neither necessary nor feasible. Instead, because the MNTES conversion function is a continuous one-dimensional function, we can pre-compute a one-dimensional lookup table for each TES. This lookup table can then be interpolated for computationally efficient power reconstruction, as done in the companion paper\cite{micheli_systematic_2024}.

\section{Conclusion}
\label{sect:conclusion}

\emph{LiteBIRD}'s mission to detect or constrain primordial B-modes with $\sigma(r)<0.001$ requires meticulous control of instrumental systematics. We have discussed ways in which nonlinearity in the TES bolometers can add introduce systematic errors to \emph{LiteBIRD} data by up-converting HWP synchronous signals to the science band, distorting the shapes of bright astrophysical sources, biasing the CMB dipole calibration, and introducing gain drifts due to drifts in the environment. To mitigate these effects, we introduced MNTES, a physics-based technique for modeling and removing TES nonlinearity through a nonlinear reconstruction of the incident optical power.

By leveraging a physics-based model of the TES detector and readout system, MNTES can correct for power-to-current conversion errors and accurately reconstruct the incident optical power over a wide dynamic range. Precise knowledge of the TES parameters and circuit properties from the on-ground calibration enables the MNTES algorithm to be applied to flight data. Importantly, the MNTES approach does not require the TES to be perfectly voltage-biased or the circuit to be purely resistive. The use of the Thévenin equivalent circuit eliminates the need to model complex bias and readout impedances in detail.

Before launch, the MNTES method will be validated on representative \emph{LiteBIRD} flight hardware in the lab. We will compare the reconstructed optical power to the known incident power for various types of input signals, including DC steps, simulated HWP modulation at $2f_\mathrm{HWP}$ and $4f_\mathrm{HWP}$, and scanning across polarized and unpolarized sources. If the MNTES technique proves as effective as anticipated, this novel approach to nonlinearity correction will ensure that \emph{LiteBIRD} reaches its full scientific potential.

\section*{Code, Data, and Materials Availability}

Data sharing is not applicable to this article, as no new data were created or analyzed. The code used in this work is available at: \href{https://github.com/tijmen/dfmux_calc/blob/main/mntes.py}{https://github.com/tijmen/dfmux\_calc}.

\acknowledgments
\emph{LiteBIRD} (phase A) activities are supported by the following funding sources: ISAS/JAXA, MEXT, JSPS, KEK (Japan); CSA (Canada); CNES, CNRS, CEA (France); DFG (Germany); ASI, INFN, INAF (Italy); RCN (Norway); MCIN/AEI, CDTI (Spain); SNSA, SRC (Sweden); UKSA (UK); and NASA, DOE (USA). Tijmen de Haan was supported by World Premier International Research Center Initiative (WPI), MEXT, Japan. 

The LiteBIRD collaboration consists of Debabrata Adak, Peter Ade, Alexandre Adler, Nabila Aghanim, Kosuke Aizawa, Hiroki Akamatsu, Yoshiki Akiba, Ryosuke Akizawa, Erwan Allys, David Alonso, Avinash Anand, Kam Arnold, Antoine Arondel, Didier Auguste, Jonathan Aumont, Ragnhild Aurlien, Jason Austermann, Susanna Azzoni, Carlo Baccigalupi, Mario Ballardini, Anthony Banday, Ranajoy Banerji, Rita Barreiro, Nicola Bartolo, Soumen Basak, Artem Basyrov, Elia Battistelli, Ludovik Bautista, Jim Beall, Dominic Beck, Shawn Beckman, Karim Benabed, Benjamin Beringue, Juan Bermejo-Ballesteros, Marco Bersanelli, Anaïs Besnard, Dmitry Blinov, Julien Bonis, Julian Borrill, Marco Bortolami, François Bouchet, Francois Boulanger, Sophie Bounissou, Charlotte Braithwaite, Maksym Brilenkov, Thejs Brinckmann, Michael Brown, Martin Bucher, Alessandro Buzzelli, Federico Cacciotti, Erminia Calabrese, Martino Calvo, Paolo Campeti, Ed Canavan, Emile Carinos, Alessandro Carones, Florie Carralot, Jerome Carron, Francisco Casas, Andrea Catalano, Giovanni Cavallotto, Abdallah Chahadih, Anthony Challinor, Victor Chan, Hiroaki Chan, Jyothis Chandran, Kolen Cheung, Meng Chiao, Yuji Chinone, Caterina Chiocchetta, Jens Chluba, Michele Citran, Susan Clark, Lionel Clermont, Sébastien Clesse, Jean-Francois Cliche, Loris Colombo, Fabio Columbro, Giulia Conenna, Gabriele Coppi, Alessandro Coppolecchia, William Coulton, Javier Cubas, Ari Cukierman, David Curtis, Francesco Cuttaia, Giuseppe D'Alessandro, Sultan Dabagov, Konstantina Dachlythra, Paolo Dal Bo, Paolo de Bernardis, Tijmen de Haan, Elena de la Hoz, Mario De Lucia, Marco De Petris, Stefano Della Torre, Giovanni Delle Monache, Karine Demyk, Eugenia Di Giorgi, José Díaz García, Clive Dickinson, Patricia Diego-Palazuelos, Jiao Ding, Matt Dobbs, Naoya Doi, Tadayasu Dotani, Denis Douillet, Eric Doumayrou, Marian Douspis, Lionel Duband, Anne Ducout, Shannon Duff, Adri Duivenvoorden, Cydalise Dumesnil, Jean-Louis Durand, Jean-Baptiste Durrive, Jean-Marc Duval, Ken Ebisawa, Tucker Elleflot, Hans Eriksen, Josquin Errard, Thomas Essinger-Hileman, Nicole Farias, Samuel Farrens, Elisa Ferreira, Katia Ferrière, Fabio Finelli, Raphael Flauger, Karl Fleury-Frenette, Cristian Franceschet, Koji Fujii, Ryuichi Fujimoto, Unni Fuskeland, Samuele Galeotta, Silvia Galli, Luca Galli, Giacomo Galloni, Mathew Galloway, Ken Ganga, Jian Gao, Thomas Gasparetto, Ricardo Génova-Santos, Marc Georges, Martina Gerbino, Massimo Gervasi, Tommaso Ghigna, Serena Giardiello, Christian Gimeno-Amo, Eirik Gjerløw, Miguel Gomes, Raúl González González, Adelie Gorce, Marcin Gradziel, Julien Grain, Laurent Grandsire, Frank Grupp, Alessandro Gruppuso, Jon Gudmundsson, Nils Halverson, Jean-Christophe Hamilton, Dariush  Hampai, Frode Hansen, Fumiya Harashima, Peter Hargrave, Stuart  Harper, Ian Harrison, Takashi Hasebe, Masaya Hasegawa, Makoto Hattori, Masashi Hazumi, Sophie Henrot-Versillé, Brandon Hensley, Lukas Hergt, Daniel Herman, Carlos Hernández-Monteagudo, Diego Herranz, Mitsuhiro Higuchi, Charles Hill, Gene Hilton, Yukimasa Hirota, Yuto Hirumi, Eric Hivon, Renee Hlozek, Thuong Hoang, Amber Hornsby, Yurika Hoshino, Johannes Hubmayr, Kiyotomo Ichiki, Teruhito Iida, Kotaro Iida, Takuro Ikemoto, Kiyoshi Ikuma, Stéphane Ilic, Hiroaki Imada, Kosuke Inoue, Kosei Ishimura, Hirokazu Ishino, Koki Ishizaka, Taisei Iwagaki, Takumi Izawa, Greg Jaehnig, Michael Jones, Baptiste Jost, Tooru Kaga, Daisuke Kaneko, Shunpei Kaneko, Shingo Kashima, Miu Kashiwazaki, Yuichiro Kataoka, Nobuhiko Katayama, Akihiro Kato, Takeo Kawasaki, Reijo Keskitalo, Atsuko Kibayashi, Kimihiro Kimura, Christian Kintziger, Theodore Kisner, Yohei Kobayashi, Nozomu Kogiso, Alan Kogut, Kazunori Kohri, Eiichiro Komatsu, Kunimoto Komatsu, Kuniaki Konishi, Nicoletta Krachmalnicoff, Joel Krajewski, Ingo Kreykenbohm, Remi Kriboo, Chao-Lin Kuo, Akihiro Kushino, Luca Lamagna, Jeff Lanen, Gregory Laquaniello, Tommaso Lari, Massimiliano Lattanzi, Rene Laureijs, Jean-Christophe Le Clec'h, Adrian Lee, Clément Leloup, Margherita Lembo, Julien Lesgourgues, François Levrier, Andrea Limonta, Eric Linder, Jia Liu, Anto Lonappan, Marcos López-Caniego, Thibaut Louis, Gemma Luzzi, Matt Lyons, Juan Macias-Perez, Thierry Maciaszek, Asuka Maeda, Stephan Maestre, Bruno Maffei, Davide Maino, Muneyoshi Maki, Stefano Mandelli, Anna Mangilli, Elenia Manzian, Vipul Prasad Maranchery, Elisabetta Marchitelli, Michele Maris, Benoit Marquet, Marshall Marshall, Enrique Martínez-González, Felice Martire, Silvia Masi, Maurizio Massa, Mika Masuzawa, Sabino Matarrese, Frederick Matsuda, Tomotake Matsumura, Mikiya Matsuoka, Lorenzo Mele, Aniello Mennella, Silvia Micheli, Marina Migliaccio, Yuto Minami, Kazuhisa Mitsuda, Fumiya Miura, Andrea Moggi, Diego Molinari, Marta Monelli, Alessandro Monfardini, Joshua Montgomery, Ludovic Montier, Gianluca Morgante, Junichi Mori, Ryosuke Moriguchi, Mami Morinaga, Baptiste Mot, Louise Mousset, Yasuhiro Murata, John Murphy, Makoto Nagai, Hosei Nagano, Yuya Nagano, Taketo Nagasaki, Ryo Nagata, Shogo Nakamura, Ryo Nakano, Toshiya Namikawa, Katsuhiro Narasaki, Federico Nati, Paolo Natoli, Simran Nerval, Toshiyuki Nishibori, Haruki Nishino, Alessandro Novelli, Fabio Noviello, Créidhe O'Sullivan, Ippei Obata, Andrea Occhiuzzi, Hiroki Ochi, Kimihide Odagiri, Hiroyuki Ogawa, Hideo Ogawa, Shugo Oguri, Hiroyuki Ohsaki, Izumi Ohta, Norio Okada, Nozomi Okada, Shunsuke Okumura, Ryuji Omae, Luca Pagano, Alessandro Paiella, Daniela Paoletti, Guillermo Pascual-Cisneros, Andrea Passerini, Guillaume Patanchon, Vasiliki Pavlidou, Vincent Pelgrims, Julien Peloton, Valeria Pettorino, Francesco Piacentini, Michel Piat, Giulia Piccirilli, Michele Pinchera, Frederic Pinsard, Giampaolo Pisano, Jean-Yves Plesseria, Gianluca Polenta, Davide Poletti, Luca Porcelli, Frederick Porter, Thomas Prouvé, Giuseppe Puglisi, Nicolò Elia Raffuzzi, Damien Rambaud, Christopher Raum, Sabrina Realini, Martin Reinecke, Mathieu Remazeilles, Isabelle Ristorcelli, Alessia Ritacco, Arianna Rizzieri, Philippe Rosier, Gilles Roudil, Jose Rubiño-Martín, Miguel Ruiz-Granda, Megan Russell, Yuki Sakurai, Haruyuki Sakurai, Laura Salvati, Maura Sandri, Joy Kaushik Sanghavi, Maria Sansa-Bernat, Manami Sasaki, Okumura Satsuki, Valentin Sauvage, Giorgio Savini, Douglas Scott, Joseph Seibert, Yutaro Sekimoto, Blake Sherwin, Keisuke Shinozaki, Maresuke Shiraishi, Peter Shirron, Alexey Shitvov, Giovanni Signorelli, Gaganpreet Singh, Graeme Smecher, Juan  Socuéllamos Chacón , Sherry Song, Franco Spinella, Jean-Luc Starck, Samantha Stever, Radek Stompor, Nils-Ole Stutzer, Rashmi Sudiwala, Hajime Sugai, Shinya Sugiyama, Raelyn Sullivan, Toyoaki Suzuki, Junichi Suzuki, Aritoki Suzuki, Trygve Svalheim, Eric Switzer, Rion Takahashi, Ryota Takaku, Hayato Takakura, Satoru Takakura, Yusuke Takase, Youichi Takeda, Hideki Tanimura, Andrea Tartari, Konstantinos Tassis, Daniele Tavagnacco, Angela Taylor, Ellen Taylor, Yutaka Terao, Luca Terenzi, Jean-Pierre Thermeau, Leander Thiele, Harald Thommesen, Keith Thompson, Ben Thorne, Takayuki Toda, Maurizio Tomasi, Hiroshi Tomida, Mayu Tominaga, Neil Trappe, Matthieu Tristram, Masatoshi Tsuji, Masahiro Tsujimoto, Carole Tucker, Ryuichi Ueki, Ryota Uematsu, Ai Ueno, Joel Ullom, Kensei Umemori, Satoru Uozumi, Shin Utsunomiya, Davide Vaccaro, Léo Vacher, Bartjan van Tent, Ian Veenendaal, Gerard Vermeulen, Patricio Vielva, Fabrizio Villa, Michael Vissers, Nicola Vittorio, Benjamin Wandelt, Wang Wang, Shengzhu (Alex) Wang, Naoki Watanabe, Kazuya Watanuki, Duncan Watts, Ingunn Wehus, Jochen Weller, Benjamin Westbrook, Gilles Weymann-Despres, Joern Wilms, Berend Winter, Edward Wollack, Ryo Yamamoto, Noriko Yamasaki, Masato Yanagisawa, Tetsuya Yoshida, Nathalie Ysard, Junji Yumoto, Andrea Zacchei, Mario Zannoni, Yu Zhou, Andrea Zonca. 

This article previously appeared in SPIE Proceedings Volume Millimeter, Submillimeter, and Far-Infrared Detectors and Instrumentation for Astronomy XII, 1310208 (2024) \href{https://doi.org/10.1117/12.3018503}{doi:10.1117/12.3018503}.

\bibliography{mntes}
\bibliographystyle{spiejour}

\end{document}